\documentstyle[aps,preprint,epsfig,floats]{revtex}

\def\slash#1{#1\!\!\!/}
\def\eqref#1{Eq.\ (\ref{#1})}

\begin{document}
\setcounter{page}{0}
\def\footnoterule{\kern-3pt \hrule width\hsize \kern3pt}
\tighten

\title{Mass-Induced Crystalline Color Superconductivity}

\author{Joydip~Kundu\footnote{Email address: {\tt kundu@mit.edu}},
Krishna~Rajagopal\footnote{Email address: {\tt krishna@ctp.mit.edu}}}

\address{Center for Theoretical Physics \\
Massachusetts Institute of Technology \\
Cambridge, MA 02139 \\
}

\date{MIT-CTP-3224, hep-ph/0112206, December 14, 2001}
\maketitle

\thispagestyle{empty}

\begin{abstract}
We demonstrate that crystalline color superconductivity 
may arise as a result of pairing between massless quarks
and quarks with nonzero mass $m_s$.
Previous analyses of this phase of cold dense quark matter
have all utilized a chemical potential difference $\delta\mu$
to favor crystalline color superconductivity over ordinary
BCS pairing. In any context in which crystalline
color superconductivity occurs in nature, however, it will be $m_s$-induced.
The effect of $m_s$ is 
qualitatively different from that of $\delta\mu$ in one crucial respect: 
$m_s$ depresses
the value of the BCS gap $\Delta_0$ whereas $\delta\mu$ leaves
$\Delta_0$ unchanged.  This effect in the BCS phase 
must be taken into account before
$m_s$-induced and $\delta\mu$-induced
crystalline color superconductivity can sensibly be compared. 
\end{abstract}

\vfill\eject

\section{Introduction}

Since quarks which are antisymmetric in color attract,
cold dense quark matter is unstable to the formation of
a condensate of Cooper pairs, making it a color 
superconductor \cite{2SC}.  
At asymptotic densities, the ground state of QCD with 
quarks of three flavors ($u$, $d$ and $s$) 
with equal masses is expected to be the color-flavor locked (CFL) 
phase \cite{CFL,OtherCFL,reviews}.
This phase features a condensate of Cooper pairs of
quarks which includes $ud$, $us$, and $ds$ pairs. Quarks
of all colors and all flavors participate in the
pairing, and all excitations with quark quantum numbers are
gapped. In this phase, left-flavor and right-flavor symmetries 
are both locked to color, breaking chiral symmetry~\cite{CFL}.
As in any BCS state, the Cooper pairing in the CFL
state pairs quarks whose momenta are equal in 
magnitude and opposite in direction, and pairing is strongest
between pairs of quarks whose momenta are both near their respective
Fermi surfaces.  Pairing persists even in the face of a stress (such
as a chemical potential difference or a mass difference)
that seeks to push the Fermi surfaces apart, although
a stress that is too 
strong will ultimately disrupt Cooper pairing~\cite{ABR2+1,SW2}.
Thus, the CFL phase persists 
for unequal quark masses, so long as the 
differences are not too large~\cite{ABR2+1,SW2}.
This means that the CFL phase is the ground
state for real QCD, assumed to be in equilibrium
with respect to the weak interactions, as long
as the density is high enough. 

Imagine decreasing the quark number chemical potential $\mu$ 
from asymptotically large values. 
The quark matter 
at first remains color-flavor locked, 
although the CFL condensate may rotate
in flavor space as terms of order $m_s^4$
in the free energy become important~\cite{BedaqueSchaefer}. 
Color-flavor locking is maintained
until a transition to a state in which 
some quarks become ungapped.  This ``unlocking transition'', which  
must be first order~\cite{ABR2+1,SW2}, occurs 
when~\cite{ABR2+1,SW2,neutrality,ARRW}
\begin{equation}\label{ApproxUnlocking}
\mu \approx m_s^2 / 4\Delta_0\ .
\end{equation}
In this expression, $\Delta_0$ is the BCS pairing gap,
estimated in both models and asymptotic analyses to be of order 
tens to 100 MeV~\cite{reviews}, and  
$m_s$ is the strange quark mass parameter.  
$m_s$ includes the contribution
from any $\langle \bar s s \rangle$ condensate 
induced by the nonzero current strange quark mass, making it a 
density-dependent effective mass, decreasing as density increases 
and equaling the current strange quark mass only at asymptotically high
densities.  
At densities which may arise at the center of compact stars, 
corresponding to $\mu\sim 400-500$ MeV, $m_s$ is certainly significantly
larger than the current quark mass, and its value is not well-known.
In fact, $m_s$ decreases discontinuously at the unlocking 
transition~\cite{BuballaOertel}.
Thus, the criterion (\ref{ApproxUnlocking}) can only be used
as a rough guide to the location of the unlocking transition
in nature~\cite{BuballaOertel}.
Given this quantitative uncertainty, there remain two logical possibilities
for what happens as a function of decreasing $\mu$.
One
possibility is a first order phase transition directly from
color-flavor locked quark matter to hadronic matter,
as explored in Ref. \cite{ARRW}.  The second possibility
is an unlocking transition~\cite{ABR2+1,SW2} to quark matter
in which not all quarks participate in the dominant pairing,
followed only at a lower $\mu$ by a transition to hadronic matter.
We assume the second possibility here, and explore its consequences.

Once CFL is disrupted, leaving some species of quarks with 
differing Fermi momenta and therefore unable to participate
in BCS pairing,  
it is natural to ask whether there is some generalization
of the ansatz in which pairing 
between two species of quarks persists even once their
Fermi momenta differ.   Crystalline color superconductivity
is the answer to this question.
The idea is that it may 
be favorable for quarks with differing Fermi momenta 
to form pairs whose momenta are {\it not} equal
in magnitude and opposite in sign~\cite{LOFF,BowersLOFF}.  
This generalization of the pairing ansatz (beyond BCS ans\"atze
in which only quarks with momenta which add to zero pair) is favored because
it gives rise to a region of phase space where {\it both}
of the quarks
in a pair are close to their respective Fermi surfaces,
and such pairs can be created at low cost in free energy.
Condensates of this sort spontaneously
break translational and rotational invariance, leading to
gaps which vary in a crystalline pattern.
As a function of increasing depth in a compact star,
$\mu$ increases, $m_s$ decreases, and $\Delta_0$ changes also.
If in some shell within the quark matter core
of a neutron star (or within a strange quark star)  
the quark number densities are
such that crystalline color superconductivity arises,
rotational vortices may be pinned in this shell, making
it a locus for glitch phenomena~\cite{BowersLOFF,LOFFreview}.

An analysis of these ideas in the context of the disruption
of CFL pairing is complicated by the fact that
in quark matter in which CFL pairing
does not occur, up and down quarks 
may nevertheless
continue to pair in the usual BCS fashion.  In this 2SC
phase, which was the earliest color superconducting 
phase to be studied~\cite{2SC},
the attractive
channel involves the formation of Cooper pairs which
are antisymmetric in both color and 
flavor, yielding
a condensate with color (Greek indices)
and flavor (Latin indices) structure 
$\langle q^\alpha_a  q^\beta_b \rangle\sim \epsilon_{ab}
\epsilon^{\alpha\beta 3}$.  
This condensate leaves five quarks unpaired: up and down quarks
of the third color, and strange quarks of all three colors.
Because the BCS pairing scheme leaves ungapped quarks with
differing Fermi momenta, crystalline color superconductivity
may result.

To date, crystalline color superconductivity has only
been studied in the simplified model context with
pairing between two quark species whose
Fermi momenta are pushed apart
by turning on a chemical
potential difference~\cite{BowersLOFF,ngloff,LOFFphonon,pertloff},
rather than considering CFL pairing in the presence of
quark mass differences.                                            
Our goal in this paper is to investigate the ways in
which the response of the system to mass differences is
similar to or different from the response to chemical potential
differences.  We can address this question within
the two-flavor model by generalizing it to describe pairing 
between massless up quarks and strange quarks with
mass $m_s$.  For completeness, we introduce
\begin{eqnarray}\label{mubardmu}
\mu_u&=&\mu-\delta\mu\nonumber\\
\mu_s&=&\mu+\delta\mu\ ,
\end{eqnarray}
allowing us to consider the effects of $m_s$ and $\delta\mu$
simultaneously.
We shall use this two-flavor toy model throughout, deferring
an analysis of crystalline color superconductivity
induced by the effects of $m_s$ on three-flavor quark
matter to future work.

\subsection{Consequences of $\delta\mu\neq 0$, with $m_s$=0}

Before describing the consequences of $m_s\neq 0$, let
us review the salient facts known about the 
consequences of $\delta\mu\neq 0$, upon taking $m_s=0$.
If $|\delta\mu|$ is nonzero but less than some $\delta\mu_1$,
the ground state in the two-flavor toy-model 
is precisely that obtained for 
$\delta\mu =0$ \cite{Clogston,Bedaque,BowersLOFF}.\footnote{In 
this two-flavor toy-model the diquark condensate
is a flavor singlet. As the condensate 
breaks no flavor symmetries, there is
no analogue of the rotations of the condensate
in flavor space which occur within the CFL 
phase with nonzero $\delta\mu$~\cite{BedaqueSchaefer,KaplanReddy}.}
In this 2SC state,
red and green up and strange quarks pair, yielding four quasiparticles
with superconducting gap $\Delta_0$ \cite{2SC}.
Furthermore, the number density of red and green up quarks is the
same as that of red and green strange quarks. 
As long as $|\delta\mu|$ is not too large, this BCS state
remains unchanged (and favored) 
because maintaining equal number densities, and
thus coincident Fermi surfaces, 
maximizes the pairing
and hence the interaction energy.  
As $|\delta\mu|$ is increased, the BCS state remains the ground state of
the system only as long as its negative interaction
energy offsets the large positive free energy cost 
associated with forcing the Fermi seas to remain coincident.
In the weak coupling limit, in
which $\Delta_0/\mu\ll 1$, the BCS state persists for
$|\delta\mu|<\delta\mu_1=\Delta_0/\sqrt{2}$ \cite{Clogston,BowersLOFF}.
For larger $\Delta_0$, the $1/\sqrt{2}$ coefficient changes in value.
These conclusions are the same 
whether the interaction between quarks is
modeled as a point-like four-fermion interaction or 
is approximated by single-gluon 
exchange~\cite{reviews}.
The loss of BCS pairing at $|\delta\mu|=\delta\mu_1$  is the
analogue in this toy model of the unlocking transition.

If $|\delta\mu|>\delta\mu_1$, BCS pairing between $u$
and $s$ is
not possible. However, in a range $\delta\mu_1<|\delta\mu|<\delta\mu_2$
near the unpairing transition,
it is favorable to form a crystalline color superconducting state in
which the Cooper pairs have nonzero momentum. 
This phenomenon was first analyzed 
by Larkin and Ovchinnikov
and Fulde and Ferrell~\cite{LOFF} (LOFF)
in the context of
pairing between electrons in which spin-up and spin-down 
Fermi momenta differ.  It has proven difficult to find a
condensed matter physics system which is well described simply 
as BCS pairing in the presence of a Zeeman effect: any 
magnetic perturbation
that may induce a Zeeman effect tends to have much 
larger effects on the motion of 
the electrons, as in the Meissner effect.  The 
QCD context of interest to us, in which the Fermi momenta
being split are those of different flavors rather than of different
spins, therefore turns out to be the
natural arena for the phenomenon first analyzed by LOFF.

The crystalline color superconducting phase (also called
the LOFF phase) has been described in 
Ref.~\cite{BowersLOFF} (following Refs. \cite{LOFF}) 
upon making the simplifying 
assumption that
quarks interact via a
four-fermion interaction
with the quantum numbers of single gluon exchange. 
In the LOFF state, each Cooper pair carries 
momentum $2{\bf q}$ with $|{\bf q}|\approx 1.2 \delta\mu$.
The condensate and gap parameter vary in space with wavelength
$\pi/|{\bf q}|$.  Although the magnitude $|{\bf q}|$ is determined
energetically, the direction $\hat{\bf q}$
is chosen spontaneously.  
The LOFF state is characterized by a gap parameter $\Delta$ and a 
diquark condensate, but not by an energy gap in the dispersion
relation: the quasiparticle dispersion 
relations vary
with the direction of the momentum, yielding gaps that vary from zero
up to a maximum of $\Delta$.  
The condensate is dominated by
those regions in momentum space in which a quark pair
with total momentum $2{\bf q}$ has both members of
the pair within $\sim \Delta$ of their respective 
Fermi surfaces. These regions form circular bands
on the two Fermi surfaces.  
Making the ansatz that all Cooper pairs make the same choice of
direction $\hat{\bf q}$ corresponds
to choosing a single circular band on each Fermi surface.
In position space, it corresponds to a condensate which
varies in space like 
\begin{equation}\label{simplifiedcondensate}
\langle \psi({\bf x}) \psi({\bf x})\rangle \propto \Delta e^{2i{\bf q}
\cdot {\bf x}}\ .  
\end{equation} 
This ansatz is certainly {\it not} the best choice.
If a single plane wave is favored, why not two? That is,
if one choice of $\hat{\bf q}$ is favored, why not add 
a second ${\bf q}$, with the same $|{\bf q}|$ but
a different $\hat{\bf q}$?  If two are favored, why not three?
This question, namely, the determination of the favored crystal
structure of the crystalline color superconductor phase,
is unresolved but is under investigation.
Note, however, that if we find a region
$\delta\mu_1<|\delta\mu|<\delta\mu_2$ in which the simple
LOFF ansatz with a single $\hat{\bf q}$ is favored over
the BCS state and over no pairing, then the LOFF state
with whatever crystal structure turns out to be optimal
must be favored in {\it at least} this region.  Note also
that even the single $\hat{\bf q}$ ansatz, which we use
henceforth, breaks translational and rotational invariance
spontaneously.  The resulting phonon has been analyzed
in Ref. \cite{LOFFphonon}.

Crystalline color superconductivity is favored within a window 
$\delta\mu_1<|\delta\mu|<\delta\mu_2$. As $|\delta\mu|$ increases from 0,
one finds a first order phase transition from the ordinary
BCS phase to the crystalline color superconducting phase
at $|\delta\mu|=\delta\mu_1$ and then a second order
phase transition at $|\delta\mu|=\delta\mu_2$ at which $\Delta$          
decreases to zero.
Because the condensation
energy in the LOFF phase is much smaller than that of the BCS condensate
at $\delta\mu=0$, the value of $\delta\mu_1$ is almost identical
to that at which the naive unpairing transition from the 
BCS state to the state with no pairing would occur if
one ignored the possibility of a LOFF phase, 
namely $\delta\mu_1=\Delta_0/\sqrt{2}$.  
For all practical
purposes, therefore, the LOFF gap equation is not required in order
to determine $\delta\mu_1$. The LOFF gap equation
is used to determine $\delta\mu_2$
and the properties of the crystalline color
superconducting phase \cite{BowersLOFF}. 
In the limit of a weak four-fermion interaction, 
the crystalline color superconductivity window is bounded by
$\delta\mu_1=\Delta_0/\sqrt{2}$ and $\delta\mu_2=0.754\Delta_0$,
as first demonstrated in Refs. \cite{LOFF}.  These results have
been extended beyond the weak four-fermion interaction 
limit in Ref. \cite{BowersLOFF}.

We now know 
that the use of the simplified point-like
interaction significantly underestimates 
the width of the LOFF window: assuming instead that quarks
interact by exchanging medium-modified gluons yields
a much larger value of $\delta\mu_2$~\cite{pertloff}.
This can be understood upon noting that quark-quark interaction
by gluon exchange is dominated by forward scattering. In
most scatterings, the angular positions on their
respective Fermi surfaces do not change much. In the LOFF state,
small-angle scattering is advantageous because it cannot
scatter a pair of quarks out of the region of momentum space
in which both members of the pair are in their respective circular
bands, where pairing is favored. This means that it is natural
that a forward-scattering dominated interaction like single-gluon
exchange is much more favorable for crystalline color 
superconductivity that a point-like interaction, which
yields $s$-wave scattering.
Thus, although for the present we shall use the 
point-like interaction in our
analysis of $m_s$-induced crystalline color superconductivity, 
it is worth remembering
that this is very conservative.

\subsection{Consequences of $m_s\neq 0$}

In the absence of any interaction, and thus in the
absence of pairing, the effect of a strange quark mass
is to shift the Fermi momenta to
\begin{eqnarray}\label{unpairedpF}
p_F^u&=&\mu-\delta\mu\nonumber\\
p_F^s&=&\sqrt{(\mu+\delta\mu)^2-m_s^2}\ .
\end{eqnarray}
Assuming both $|\delta\mu|/\mu$ and $m_s/\mu$ are small, the separation
between the two Fermi momenta is $\approx |2\delta\mu - m_s^2/2\mu|$.
This suggests the conjecture that even when $m_s\neq 0$ the description
given in the previous subsection continues to be valid upon
replacing $|\delta\mu|$ by $|\delta\mu - m_s^2/4\mu|$.  We shall
see in Section 3
that this conjecture is {\it incorrect} in one key respect:
whereas if $m_s=0$ a $|\delta\mu|$ which is nonzero but smaller
than $\delta\mu_1$ has no effect on the BCS state, the BCS gap $\Delta_0$
decreases with increasing $m_s^2$, as we discuss
further in an appendix.
We show that for small $m_s^2$,
$\Delta_0(m_s)/\Delta_0(0)$ decreases linearly 
with $m_s^2$.
Because $\Delta_0$ occurs in the free energy in a term of
order $\Delta_0^2\mu^2$, the $m_s$-dependence of $\Delta_0$
corrects the free energy by of order $\Delta_0(0)^2 m_s^2$.
As $\delta\mu$ has no analogous effect, we conclude that
$m_s^2/4\mu$ and $\delta\mu$ have qualitatively
different effects on the paired state.

At another level, however, the story {\it is} quite
similar to that for $m_s=0$: if $|\delta\mu - m_s^2/4\mu|$
is small enough, we find the BCS state; if $|\delta\mu - m_s^2/4\mu|$
lies within an intermediate window, we find LOFF pairing;
if $|\delta\mu - m_s^2/4\mu|$ is large enough, no pairing is possible.
The boundaries between the phases, however, are related to
a $\Delta_0(m_s)$, rather than simply to a constant $\Delta_0$.
That is, the definitions of ``small enough'' and ``large enough''
are $m_s$-dependent.
We map the $(m_s,\delta\mu)$ plane in this paper. In Section 2,
we derive a gap equation which we then use in Section 3
to analyze the BCS phase, finding the region of the $(m_s,\delta\mu)$ 
plane in which BCS pairing is favored, and again use in Section 4 to 
analyze crystalline color superconductivity, finding windows in the 
$(m_s,\delta\mu)$ plane in which LOFF pairing is favored.

\section{The Gap Equation}

In this Section, we sketch the derivation of the gap equation
which describes either the BCS or the LOFF states in 
a two-flavor theory with $m_s\neq 0$ and $\delta\mu \neq 0$.
In the crystalline color superconducting phase, the condensate
contains pairs of $u$ and $s$ quarks with momenta such that
the total momentum of each Cooper pair is given by $2{\bf q}$,
with the direction of $\bf q$ chosen spontaneously. As noted
above, wherever there is an instability towards (\ref{simplifiedcondensate}),
we expect the true ground state to be a crystalline
condensate which varies in space like a sum of several such
plane waves with the same $|{\bf q}|$.  (To recover the BCS
gap equation from that which we now derive, simply set $|{\bf q}|=0$.)

In order to describe pairing between $u$ quarks 
with momentum $\bf{p+q}$ and $s$ quarks with momentum $\bf{-p+q}$, we 
use a modified Nambu-Gorkov propagator:
\begin{equation}\label{Psi}
\Psi(p,q)= \left(\begin{array}{l} \psi_u(p+q) \\ \psi_s(p-q) \\ 
\bar{\psi}_s^T (-p+q) \\ \bar{\psi}_u^T(-p-q) \end{array} \right)\ .
\end{equation}
Note that by $q$ we mean the four-vector $(0,{\bf q})$.
The Cooper pairs have nonzero total momentum but the ground
state condensate (\ref{simplifiedcondensate}) is static.
The momentum dependence of (\ref{Psi})
is motivated by the fact that in the presence of
a crystalline color superconducting condensate,
anomalous propagation does not only mean picking up or
losing two quarks from the condensate. It also means picking
up or losing momentum $2{\bf q}$.
The basis (\ref{Psi}) has been chosen so that the 
inverse fermion propagator in the crystalline
color superconducting phase is diagonal in $p$-space
and is given by
\begin{equation}\label{Sinv}
S^{-1}(p,q) = \left(\begin{array}{cccc} \slash{p}+\slash{q}+\mu_u \gamma_0
& 0 & -\bar{\bf \Delta}(p,-q) & 0 \\ 0 & \slash{p}-\slash{q}+\mu_s \gamma_0 - 
m_s& 0 & \bar{\bf \Delta}(p,q) \\ -{\bf \Delta}(p,-q) & 0 &
(\slash{p}-\slash{q}-\mu_s \gamma_0 + m_s)^T & 0 \\ 0 & {\bf \Delta}(p,q) 
& 0 & (\slash{p}+\slash{q}-\mu_u \gamma_0)^T \end{array}\right)
\end{equation}
where $\bar{\bf \Delta} = \gamma_0 {\bf \Delta}^{\dagger} \gamma_0$
and the matrix ${\bf \Delta}$ is given by
${\bf \Delta}=\Delta C\gamma_5\epsilon^{\alpha\beta 3}$. 
Note that the condensate
is explicitly antisymmetric in flavor.
$2{\bf p}$ is the relative momentum
of the quarks in a given pair and is different for different
pairs. In the gap equation below, we shall integrate over $p_0$
and ${\bf p}$. As desired, the off-diagonal blocks describe
anomalous propagation in the presence of a condensate of
diquarks with momentum $2{\bf q}$. 

We obtain the gap equation by solving
the one-loop Schwinger-Dyson equation given by
\begin{equation} \label{SDeq}
 S^{-1}(k,q)-S_0^{-1}(k,q) = -g^2 \int \frac{d^4p}{(2\pi)^4}
 \Gamma_\mu^A S(p,q)\Gamma_\nu^B D_{AB}^{\mu\nu}(k-p),
\end{equation}
where  $D_{AB}^{\mu\nu} = D^{\mu\nu} \delta_{AB}$ is the gluon
propagator, $S$ is the full quark propagator, whose inverse is 
given by (\ref{Sinv}), and $S_0$ is the fermion
propagator in the absence of interaction, given by $S$ with
${\Delta}=0$. The
vertex  is defined as 
\begin{equation} \label{vertex}
\Gamma_\mu^A = \left(\begin{array}{cccc} \gamma_\mu\lambda^A/2 & 0 & 0
& 0 \\ 0 & \gamma_\mu\lambda^A/2 & 0 & 0 \\ 0 & 0 &
-(\gamma_\mu\lambda^A/2)^T & 0 \\ 0 & 0 & 0 &
-(\gamma_\mu\lambda^A/2)^T \end{array}\right) .
\end{equation}

As in Refs. \cite{BowersLOFF,ngloff}, we simplify the interaction
between quarks by making it point-like.  That is, we replace
$g^2 D^{\mu\nu}$ by $3G g^{\mu\nu}$.  Our 
model Hamiltonian has two parameters, the four-fermion 
coupling $G$ and a cutoff $\Lambda$ on the three-momentum
${\bf p}$.  
The dependence on the cutoff
has been studied in Ref. \cite{BowersLOFF}:
upon varying $\Lambda$
while at the same time varying the coupling $G$ such
that $\Delta_0$ is kept fixed, the relation
between other physical quantities and $\Delta_0$ is
reasonably insensitive to variation of $\Lambda$.
We 
often use the value of the BCS gap $\Delta_0$
obtained at $m_s=\delta\mu=0$ to 
parametrize the strength of the interaction.
We can do this because $\Delta_0$ increases monotonically
with increasing $G$. Doing so proves convenient because
both $\delta\mu_1$ and $\delta\mu_2$ are given simply
when written in terms of the physical quantity $\Delta_0$,
whereas writing them in terms of the model-dependent
parameters $G$ and $\Lambda$ requires unwieldy expressions.

After some algebra (essentially the determination of $S$ given
$S^{-1}$ specified above), and upon suitable projection,
the Schwinger-Dyson equation (\ref{SDeq}) reduces
to a gap equation for the gap parameter $\Delta$ given
(in Euclidean space) by
\begin{equation} \label{gapeq}
\Delta = \int \frac{d^4p}{(2 \pi)^4} \frac{8 G \Delta 
(\Delta^2 + w^2 )}% p^2 + {\mu}^2 - q^2 +(p_0+i \delta\mu)^2)}
{\Delta^4+2 \Delta^2 w^2 
%(p^2 + {\mu}^2 -q^2 +(p_0+i \delta\mu)^2) 
+ \left[(p_0+i{\mu}+i\delta\mu)^2 - |{\bf p}-{\bf q}|^2 - m_s^2\right]
\left[(p_0-i{\mu}+i\delta\mu)^2 - |{\bf p}+{\bf q}|^2\right]}\ ,
\end{equation}
where 
\begin{equation}
w^2 \equiv p^2 + {\mu}^2 - q^2 +(p_0+i \delta\mu)^2\ .
\end{equation}
Eq. (\ref{gapeq}) reduces to the LOFF
gap equation of Ref.~\cite{ngloff} when $m_s=0$.  
The gap equation (\ref{gapeq}) includes the contributions
of pairing involving antiparticles, obviously
far from the Fermi surfaces. These contributions can
be removed using the methods of Ref.~\cite{ngloff} 
(and if one does so and then sets $m_s=0$ one obtains
the gap equation of Ref.~\cite{BowersLOFF}).
However, as the antiparticle contributions are suppressed
by $\Delta/\mu$ and as removing them
leads to rather unwieldy expressions, we shall simply use eq. (\ref{gapeq}).

\section{The BCS Gap and the BCS Region}

\begin{figure}[t]
\centerline{\epsfxsize 11cm \epsffile{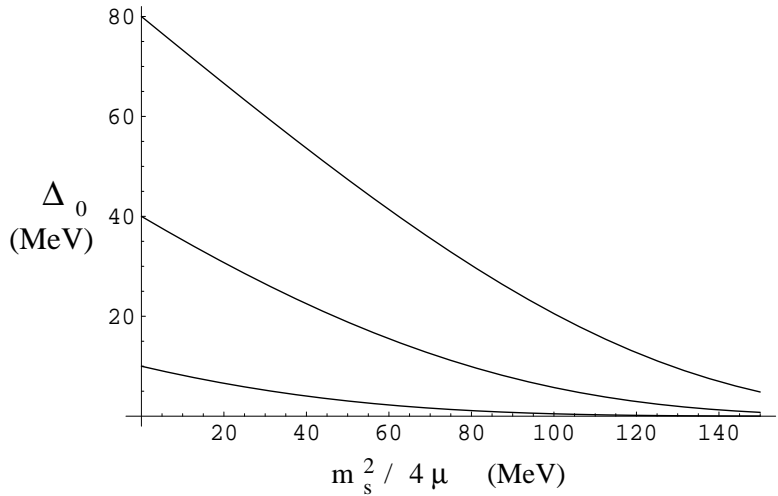}}
\caption{BCS gap $\Delta_0$ vs. $m_s^2/4\mu$, with $\delta\mu$
set equal to $m_s^2/4\mu$ to ensure that in the absence of
any pairing, the $u$ and $s$ Fermi momenta would be equal.
In this and in all subsequent
figures, we set $\mu=400$~MeV and choose a cutoff $\Lambda=1000$~MeV. 
We have chosen
three different values of the 
coupling constant $G$, yielding $\Delta_0=10,40,80$~MeV at
$m_s=0$.  $\Delta_0$ decreases with increasing
$m_s^2$, linearly for small $m_s^2$.} 
\label{Delta0vsmsFig}
\end{figure}

We expect BCS pairing to be possible when the $u$ and $s$
Fermi momenta in noninteracting quark matter are similar in value.
From (\ref{unpairedpF}) we see that $p_F^u=p_F^s$ in noninteracting
quark matter when
\begin{equation}\label{EqualpF}
\delta\mu=\frac{m_s^2}{4\mu}\ .
\end{equation}
Note that we have {\it not} 
assumed that $\delta\mu$ or $m_s$
is small compared to $\mu$ in this expression.
Let us now analyze the BCS gap equation, keeping
this in mind. Upon setting ${\bf q} = 0$,
Eq. (\ref{gapeq}) reduces to the BCS gap equation of Ref.~\cite{neutrality}: 
\begin{equation} \label{bcsgapeq}
\Delta_0 = \int \frac{d^4p}{(2 \pi)^4} \frac{8 G\Delta_0 
(\Delta_0^2 + w_0^2 )}
{\Delta_0^4+2 \Delta_0^2 w_0^2 
+ \left[(p_0+i{\mu}+i\delta\mu)^2 - p^2 - m_s^2\right]
\left[(p_0-i{\mu}+i\delta\mu)^2 - p^2\right]}\ ,
\end{equation}
where
$$w_0^2 = p^2 + {\mu}^2 + (p_0 +i \delta\mu)^2$$
and where we use the notation $\Delta_0$ for the BCS gap, to
keep track of the fact that here ${\bf q} = 0$.
Eq. (\ref{bcsgapeq}) involves an integration over $p_0$ from $-\infty$ 
to $+\infty$, which can be thought of as a contour in the complex 
$p_0$-plane, which we choose to close in the upper half-plane.
With ${\bf q}=0$, the poles in the integrand of Eq. (\ref{bcsgapeq}) 
are easily analyzed.  There are
four poles in the $p_0$-plane.  
Of these, two correspond to the contribution of antiparticle
pairing and have residues which are suppressed by $\Delta_0/\mu$
relative to the residues of the two remaining poles,
which correspond to particle pairing. The residues of the two
particle poles are equal in magnitude and opposite in sign.
As a result, 
the existence of a nonzero $\Delta_0$ which solves (\ref{bcsgapeq})
requires that one and only one of the particle poles lies
in the upper half-plane.
If we choose $\delta\mu$ according to (\ref{EqualpF}), we do
in fact find one particle pole in the upper half-plane
and thus a nonzero, $m_s$-dependent, solution $\Delta_0$ 
as shown in Fig.~\ref{Delta0vsmsFig}.
$\Delta_0$ decreases with $m_s^2$, linearly
for small $m_s^2$.
The $m_s$-dependence of $\Delta_0$ can be further understood,
but as this would divert us we defer doing so to an appendix.

What is the effect of changing $\delta\mu$ away
from the choice (\ref{EqualpF})?
This is simply understood 
upon noticing that $p_0$ and 
$\delta\mu$ enter the 
gap equation only in the combination $p_0+i\delta\mu$~\cite{neutrality}.
This means that the only effect of $\delta\mu$ is to shift
the location of the poles. Whereas $m_s$ affects the value
of the residues, $\delta\mu$ does not. 
As long as $|\delta\mu-m_s^2/4\mu|\lesssim\Delta_0$,\footnote{The 
criterion is $|\delta\mu-m_s^2/4\mu|<\Delta_0$
in the limit in which all these quantities are small
compared to $\mu$. As we shall see momentarily, this is
not the most important criterion anyway, and we therefore do
not specify it precisely.}
there is one particle pole in the upper half-plane 
and $\Delta_0$ is identical to that found when 
$\delta\mu=m_s^2/4\mu$, as in Fig.~\ref{Delta0vsmsFig}.
Once $|\delta\mu-m_s^2/4\mu|\gtrsim\Delta_0$,
there are either zero or two particle poles in the upper half-plane
and therefore no such solutions $\Delta_0$ exist.

\begin{figure}[t]
\centerline{\epsfxsize 11cm \epsffile{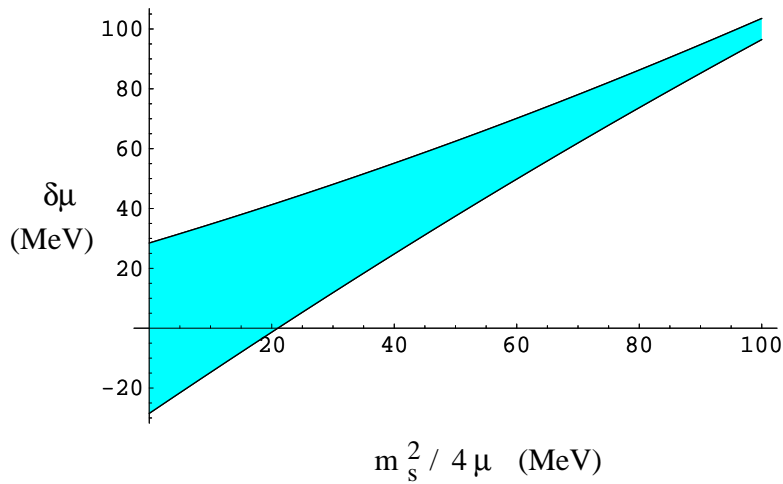}}
\caption{The region in the $(m_s^2/4\mu,\delta\mu)$  plane
where the BCS state is favored over unpaired quark matter.
We have chosen a single value of the coupling, corresponding
to $\Delta_0(0)=40$~MeV. At each value of $m_s^2/4\mu$, $\Delta_0(m_s)$
is independent of $\delta\mu$ within the shaded region
and zero outside it. $\Delta_0(m_s)$ decreases with $m_s^2/4\mu$ as 
shown in Fig.~\protect\ref{Delta0vsmsFig}.}
\label{BCSregionFig}
\end{figure}

As long as $|\delta\mu-m_s^2/4\mu|\lesssim\Delta_0(m_s)$, we have 
found a $\delta\mu$-independent, $m_s$-dependent solution $\Delta_0(m_s)$
(shown in Fig.~\ref{Delta0vsmsFig}) 
that describes a BCS state which is a
local minimum of the free energy $\Omega$ at a given
$m_s$ and $\delta\mu$.  
However, this BCS
state need not be global minimum of the free energy.
To determine the region in the $(m_s^2/4\mu,\delta\mu)$ plane
in which BCS pairing is favored over unpaired quark
matter, we must evaluate the free energy 
difference between the BCS state and the unpaired state. 
This is given by integrating the gap equation over $\Delta$ up to the 
equation's solution, $\Delta_0(m_s)$.
\begin{equation}
\Omega_{\rm BCS} - \Omega_{\rm unpaired} = \int_0^{\Delta_0(m_s)}d\Delta 
\left(\frac{\Delta}{G} - \int \frac{d^4 p}{(2 \pi)^4}\, {\rm integrand} 
\right),
\end{equation}
where ``integrand'' refers to that of the right-hand side of Eq. 
(\ref{bcsgapeq}).  The curves in the $(m_s^2/4\mu,\delta\mu)$ plane along 
which the above expression vanishes determine the region in which
the BCS phase is favored.
These curves are plotted in Fig.~\ref{BCSregionFig}. 
At any given $m_s$, BCS pairing is favored within a range
of values of $\delta\mu$ around $m_s^2/4\mu$.
Everywhere within this range, $\Delta_0$ takes on the
same ($m_s$-dependent; $\delta\mu$-independent) value.
The boundaries of the BCS region are given approximately by
\begin{equation}
\left| \delta\mu-\frac{m_s^2}{4 \mu}\right| < \frac{\Delta_0(m_s)}{\sqrt{2}}\ ,
\end{equation}
derived in Ref. \cite{neutrality} and 
valid if $\Delta_0(m_s)$, $m_s$ and $\delta\mu$ are all
small compared to $\mu$. 
We see that as $m_s$ increases and $\Delta_0(m_s)$ decreases,
the BCS region narrows.  

\vfill\eject

\section{The Crystalline Color Superconductivity Windows}

We now determine the regions of the $(m_s^2/4\mu,\delta\mu)$ plane
where the crystalline color superconducting phase, with 
$|{\bf q}|\neq 0$, is favored. These regions will be
strips just above and just below the BCS region of   
Fig.~\ref{BCSregionFig}.
Let us denote
the upper boundary of the BCS 
region $\delta\mu_1(m_s)$.  We now show that 
at a given $m_s$, the crystalline color superconducting
state is favored in some region 
$\delta\mu_1(m_s)<\delta\mu<\delta\mu_2(m_s)$.\footnote{The 
value of $\delta\mu_1(m_s)$ was determined in the previous
section by comparing the free energy of the BCS and {\it unpaired}
states. Here, we should in principle compare BCS and LOFF states
to determine $\delta\mu_1(m_s)$, but this makes very little
difference~\cite{BowersLOFF}.}
There is also a crystalline color superconductivity
window of $m_s$-dependent width just below the BCS region.

\begin{figure}[t]
\centerline{\epsfxsize 11cm \epsffile{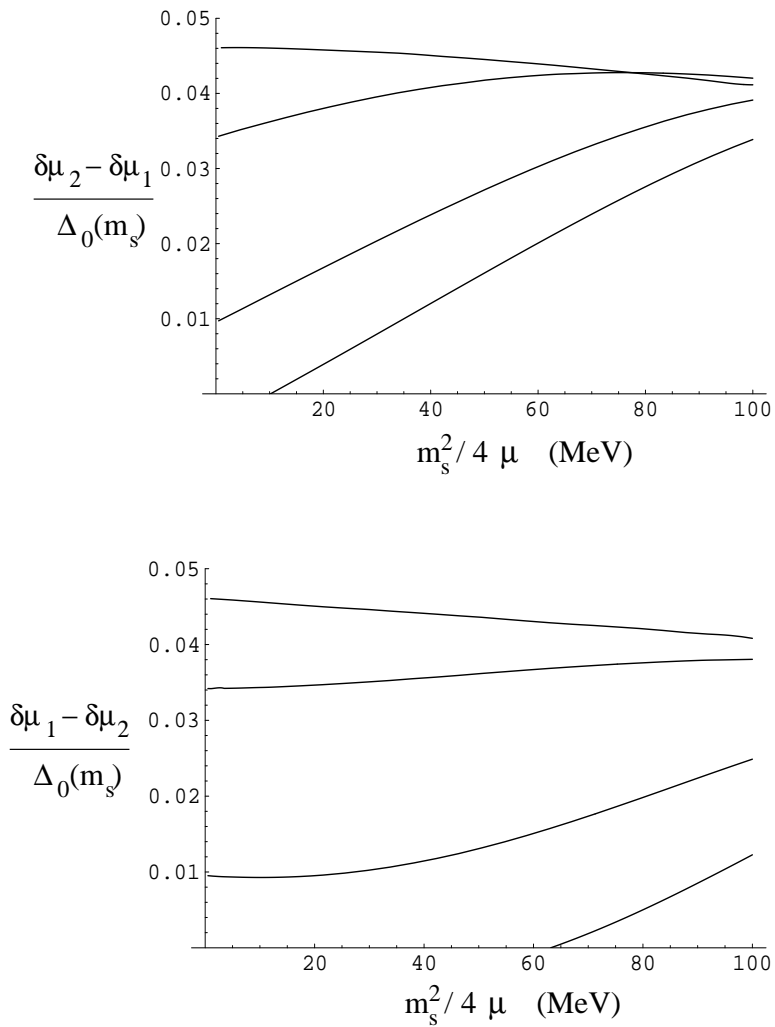}}
\caption{Upper panel: width
of the crystalline color superconductivity window above the BCS 
region. At a given $m_s$,
crystalline color superconductivity occurs for 
$\delta\mu_1(m_s)<\delta\mu<\delta\mu_2(m_s)$.
Here, we show the width of the 
window $\delta\mu_2(m_s)-\delta\mu_1(m_s)$ relative to
$\Delta_0(m_s)$ for four values of the coupling.
The four curves from top to bottom correspond  
to couplings for which 
$\Delta_0(0)=10,40,80,100$~MeV. Lower panel: width of the 
crystalline color superconductivity window below the BCS region.
The curves from top to bottom correspond to
increasing couplings, as in the upper panel.
}  
\label{punchlineFig}
\end{figure}
At $\delta\mu=\delta\mu_2$, the gap $\Delta$ in the 
LOFF state goes continuously to zero.   This
means that we can determine $\delta\mu_2$ as follows.
We set
$\Delta=0$ in the gap equation (\ref{gapeq}),
obtaining a ``zero-gap relation''  among 
$\delta\mu$, $m_s$, and $|{\bf q}|$.
We choose a value of $m_s$, and find the largest value of 
$\delta\mu$ at which the zero-gap relation can be satisfied.
This is $\delta\mu_2$.  (The smallest value of $\delta\mu$
at which the zero-gap relation can be satisfied is the lower
boundary of the crystalline color superconductivity window below
the BCS region of Fig.~\ref{BCSregionFig}.)
In Fig.~\ref{punchlineFig} we plot 
$[\delta\mu_2(m_s)-\delta\mu_1(m_s)]/\Delta_0(m_s)$
vs. $m_s^2/4\mu$ for four different values of the coupling,
corresponding to $\Delta_0(0)=10,40,80,100$~MeV.
At $m_s=0$ and in the weak coupling limit in which
$\Delta_0(0)\rightarrow 0$, 
$[\delta\mu_2(0)-\delta\mu_1(0)]/\Delta_0(0) \rightarrow
[0.754-0.707]=0.047$~\cite{LOFF}.
The results at $m_s=0$ with $\Delta_0\neq 0$ which we find in
Fig.~\ref{punchlineFig} agree with those
derived
in Ref.~\cite{BowersLOFF}. These results are symmetric about
$\delta\mu=0$.  For $m_s\neq 0$, 
we now see that when it is scaled by $\Delta_0(m_s)$,
the width of the crystalline color superconductivity
window changes little, for smaller values of $\Delta_0(0)$.  For larger 
values of $\Delta_0(0)$, the width of the window 
increases as $m_s$ increases from zero.  Indeed, we see that for 
$\Delta_0(0)=100$~MeV, the window has closed at $m_s=0$
whereas crystalline color superconductivity continues to occur at
large enough $m_s$. 

What do we conclude from our results?  First, it
is clear that the appropriate variable to use to describe
the width of the crystalline color superconductivity
window is $\delta\mu/\Delta_0(m_s)$, as opposed to $\delta\mu/\Delta_0(0)$.
When plotted as in Fig.~\ref{punchlineFig}, the width
of the window is almost independent of $m_s$ at weak coupling,
indicating that we have found the correct scaling variable.
This means that if we were to plot the crystalline
color superconducting windows as strips above and below
the BCS region of Fig.~\ref{BCSregionFig},
the ``horizontal width'' of the window on 
the $\delta\mu=0$ axis, expressed as a window in 
$m_s^2/[4\mu\Delta_0(m_s)]$, is the same as the ``vertical width''
of the
windows in $\delta\mu/\Delta_0(0)$ on the $m_s=0$ axis.
Thus, to the extent that the curves in  Fig.~\ref{punchlineFig}
are horizontal we conclude that $m_s$-induced and $\delta\mu$-induced
crystalline color superconductivity are equally robust. 
However, the curves in  Fig.~\ref{punchlineFig} are 
not quite horizontal.  At all but the weakest of couplings,
the width of the crystalline color superconductivity window
increases with $m_s$, meaning that
crystalline color superconductivity is somewhat
more robust if it is $m_s$-induced than if it is $\delta\mu$-induced.
Indeed, we find that at the moderate coupling corresponding
to $\Delta_0(0)=100$~MeV,  $m_s$-induced crystalline color superconductivity
occurs whereas $\delta\mu$-induced crystalline color superconductivity
does not.

\acknowledgments
We acknowledge helpful conversations with M. Alford, J. Bowers, 
B. Fore, A. Leibovich and E. Shuster.

Research supported in part by the DOE under cooperative research agreement
DE-FC02-94ER40818.  The work of JK is supported in part by a DOD National 
Defense Science and Engineering Graduate Fellowship.

\vfill\eject

\appendix
\section{The $m_s$-dependence of $\Delta_0$}

\begin{figure}[t]
\centerline{\epsfxsize 11cm \epsffile{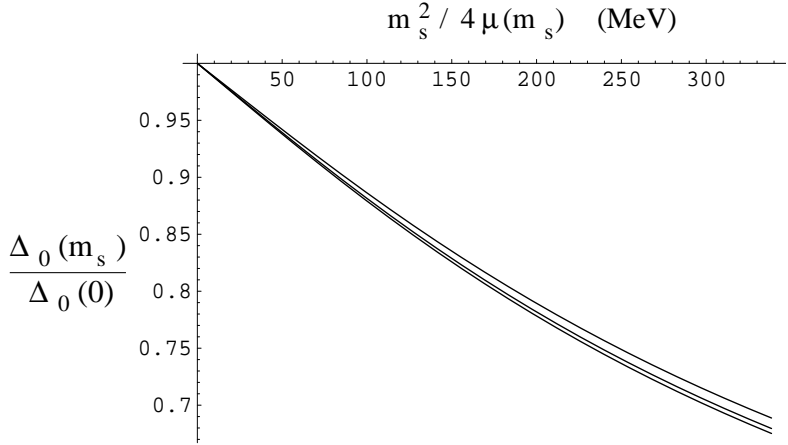}}
\caption{BCS gap $\Delta_0$ vs. $m_s^2/4\mu(m_s)$, with 
$\mu(m_s)$  
taken to increase with increasing $m_s$ in just such
a way that, in the absence of interaction, the Fermi momenta
of both the $u$ and $s$ quarks are $m_s$-independent.  
We plot the ratio $\Delta_0(m_s)/\Delta_0(0)$,
for couplings corresponding
to $\Delta_0(0)=10,40,80$~MeV, from bottom
to top. The curves are essentially independent
of the coupling when plotted in this way.
}
\label{appendixFig}
\end{figure}
In this appendix, we seek to better understand the $m_s$-dependence
of the BCS gap $\Delta_0$, shown in Fig.~\ref{Delta0vsmsFig}.
To the extent that the coupling can be taken as weak when
$\Delta_0=80$~MeV, one might expect that when each of these
three curves are scaled by their value of $\Delta_0(0)$,
they should fall on top of one another. This does not occur.
The reason for this can be understood upon first setting the interaction
to zero.  In the absence of pairing, increasing $m_s$
while setting $\delta\mu=m_s^2/4\mu$ to ensure
that $p_F^u=p_F^s$ has two effects: the value of $p_F$ 
decreases, and the strange quark Fermi velocity decreases.
Since two different effects are at play, it is not surprising
that the curves in Fig.~\ref{Delta0vsmsFig} do not scale trivially.
To disentangle these effects, we have redone the analysis
of Fig.~\ref{Delta0vsmsFig} as follows. 
As we increase $m_s$,
we now increase $\mu$, while keeping $\delta\mu=m_s^2/4\mu(m_s)$. 
As before, the latter condition ensures that, in the absence of
interaction, $p_F^u=p_F^s$. Here, we choose 
$\mu(m_s)$ in just such a way that both $p_F^u$ and $p_F^s$ 
are $m_s$-independent.
Any remaining $m_s$-dependence of $\Delta_0$ can now be seen
as arising from the variation of the difference between
the Fermi velocities in the unpaired state. The plot
analogous to Fig.~\ref{Delta0vsmsFig} is given in 
Fig.~\ref{appendixFig}, and the scaling is
now manifest: the curves are almost independent of coupling.   
Note also that $\Delta_0$ decreases linearly with $m_s^2$
as before, but the slope in Fig.~\ref{appendixFig} is much
less than that in Fig.~\ref{Delta0vsmsFig}.  This indicates
that the $m_s$-dependence in Fig.~\ref{Delta0vsmsFig} can 
be attributed 
more to the variation of $p_F$ in the unpaired state than
to the variation of $v_F$ in the unpaired state.

\end{document}